\documentclass[
    ,final            
  ]
  {aipproc}

\layoutstyle{6x9}

\begin{document}

\title{Astrophysical Reaction Rates as a Challenge for Nuclear Reaction Theory}


\author{T. Rauscher}{
  address={Department of Physics, University of Basel, CH-4056 Basel, Switzerland}
}

\begin{abstract}
The relevant energy ranges for stellar nuclear reactions are introduced. Low-energy compound and direct reactions are discussed. Stellar modifications of the cross sections are presented. Implications for experiments are outlined.
\end{abstract}

\maketitle


\section{Introduction}

Astrophysical reaction rates describe the number of reactions in a given volume of the stellar plasma and are the essential ingredient for calculations following nucleosynthesis and energy generation in astrophysical environments. They are derived by folding experimental or theoretical reaction cross sections with the energy distribution appropriate for the interacting plasma constituents. For nucleons and nuclei in stars this is the Maxwell-Boltzmann distribution (MBD). A direct determination of the cross sections faces several problems. The majority of nuclei participating in explosive nucleosynthesis is short-lived and thus the reactions cannot (yet) be studied in the laboratory. Thermonuclear reactions on stable and unstable targets proceed at low, often subCoulomb, energies, resulting in tiny reaction cross sections which are further obstacles to experiments. On the theoretical side, several reaction mechanisms may contribute, each with its own peculiar uncertainties. This makes the determination of reaction rates at astrophysically relevant energies a complex puzzle which currently can only be tackled by using several different approaches. In the following I briefly outline a few aspects important for all investigations.

\section{Compound reactions}

The majority of astrophysically relevant reactions proceeds through a compound nucleus (CN). We can distinguish between compound reactions exhibiting pronounced resonance features in their excitation functions and such without those features. The nuclear level density (NLD) at the compound formation energy (i.e. projectile energy plus separation energy of the projectile in the CN) is the distinguishing factor. A high NLD results in a large number of tightly overlapping resonances which cannot be resolved whereas a low NLD leads to more or less isolated resonance structures. Systems with a high NLD can be treated in a statistical approach, assuming an average over the NLD. This is called the Hauser-Feshbach model (HFM). Isolated resonances (and their interference) have to be treated in phenomenological approaches such as the Breit-Wigner formula (BWF) or R-matrix fits.

\subsection{Transmission coefficients and widths}

The reaction cross section of the CN reaction $a+A \rightarrow b+B$ is defined by summing over strength functions, widths or transmission coefficients
\begin{equation}
\label{eq:cs}
\sigma \propto \sum_n (2J_n+1) \frac{^n\Gamma'_a\, ^n\Gamma_b}{^n\Gamma_\mathrm{tot}}\,,
\end{equation}
where the sum is over $m$ individual resonances (BWF, $n=1\dots m$) or over all spin and parity pairs in the compound system at its formation energy (HFM). In the BWF the $\Gamma$s in the numerator are the partial widths of the resonances in the entrance and exit channels. In the HFM these are \textit{averaged} widths $\Gamma=\langle\Gamma\rangle$ which are related to strength functions $\mathcal{S}=\rho \langle\Gamma\rangle$ and transmission coefficients $T=2\pi \rho \langle\Gamma\rangle=2\pi \mathcal{S}$. The density of levels with given spin $J$ and parity $\pi$ in the compound system at the formation energy is denoted by $\rho$. In both cases, $\Gamma_\mathrm{tot}$ is the total width computed from the sum of all partial or averaged widths for all open reaction channels (including re-emission of the projectile). It is noteworthy that $\Gamma_b$ (as well as $\Gamma_\mathrm{tot}$) involve sums over transitions to all possible final states in the given channel whereas $\Gamma'_a$ only includes the transitions from the target ground state (g.s.) when laboratory cross sections are to be calculated. Additionally, all widths for a given transition include sums over the allowed partial waves.

Equation \eqref{eq:cs} and the (averaged) widths entering are essential to understanding a number of intricacies connected to the theoretical and experimental determination of cross sections for astrophysical reaction rates. This is outlined in a few examples in the following chapters.

\subsection{Relevant energy ranges}

Nuclear reactions in stellar plasmas proceed at interaction energies which are low by nuclear physics standards, even in explosive nucleosynthesis. The energy range contributing most to a stellar rate $r$ at given stellar temperature $T$ is found by inspection of the integral arising from folding the reaction cross sections with the MBD
\begin{equation}
r \propto \frac{1}{T^{3/2}}\int_0 ^\infty \sigma(E) E e^{-E/(kT)} \, dE\,. \label{eq:integral}
\end{equation}
Although the integration formally runs from Zero to Infinity most of the contributions to the integral come from a narrowly defined energy region $E_0 \pm (\Delta E)/2$. For charged particles this relevant energy region is usually called the Gamow window. The relevant energy region is determined by the energy dependences of the cross sections and the MBD. Frequently used are simple approximation formulas for the Gamow window (e.g., as given in \cite{cauldrons,ilibook}). These are based on the implicit assumption that $\Gamma'_a \ll \Gamma_b$. It has been pointed out that this assumption is not always fulfilled, especially for intermediate and heavy target nuclei \cite{ilibook,newt,tommyranges}. Using Eq.\ \eqref{eq:cs} it is easy to see why. If the largest width in the numerator is also dominating $\Gamma_\mathrm{tot}$ it will cancel with the denominator and the energy dependence will be solely given by the smaller width in the numerator. If this happens to be the entrance (projectile) width, the application of the standard approximation formulas is justified. However, in many cases it is not.

A straightforward example is the one of capture. At sizeable projectile energies, the $\gamma$ width will always be smaller than the projectile width. This will certainly apply to neutron capture. Fortunately, the energy dependence of s-, p-, and d-waves is much weaker than the exponential behavior of the MBD and in consequence the relevant energy window is given by the peak of the MBD. The fact that the $\gamma$ width is the smallest also holds for charged-particle capture, unless with light targets or at low interaction energies. In the latter case, the projectile width can be suppressed by the Coulomb barrier and may become smaller than the photon width. This depends on the structure of the nucleus and the reaction energetics and has to be scrutinized separately for each reaction. Another complication is to have different Coulomb barriers in the entrance and the exit channel. Which width is the smallest will then sensitively depend on the projectile energy and the reaction $Q$ value.

A detailed investigation of these issues and newly derived ranges of astrophysically relevant energies can be found in \cite{tommyranges}. Shifts by several MeV of the energy windows have been found. It is strongly recommended to consider these energy windows instead of the ones calculated with the standard approximation formulas, especially when devising experiments of astrophysical interest.

\section{Experimental considerations}
\subsection{Determination of astrophysically relevant cross sections}

Equation \eqref{eq:cs} is also central to the question of how much astrophysically relevant information can be extracted from a measurement of reaction cross sections. Foremostly it has to be attempted to fully cover the relevant energy window with measurements. The energy windows given in \cite{tommyranges} are defined in such a way that knowledge of the cross sections across the full energy range will determine the reaction rate integral (Eq.\ \ref{eq:integral}) to 10\%. For higher accuracy, measurements have to extend to higher and \textit{lower} energies.

Frequently, it is not possible to fully cover the relevant energy range. Especially for charged-particle reactions at intermediate and heavy targets, often it is not even possible to reach the Gamow window because subCoulomb reactions exhibit tiny cross sections. In this case measurements at higher energy are extrapolated towards lower energy or are compared to calculations to check the theoretical models. Caution is advised in these attempts because the energy dependence of the cross section and the sensitivity of the calculation to certain nuclear inputs can change depending on the energy under investigation. Again, Eq.\ \eqref{eq:cs} provides guidance. Energy dependence and sensitivity are determined by the smallest width in the numerator. (Additional complications arise when both widths are comparable or another channel is considerably contributing to $\Gamma_\mathrm{tot}$.) Since the sizes of the widths relative to each other change with energy, the smallest width may be found in another channel at low energy than at high energy.

\begin{figure}
  \includegraphics[angle=-90,width=0.8\columnwidth]{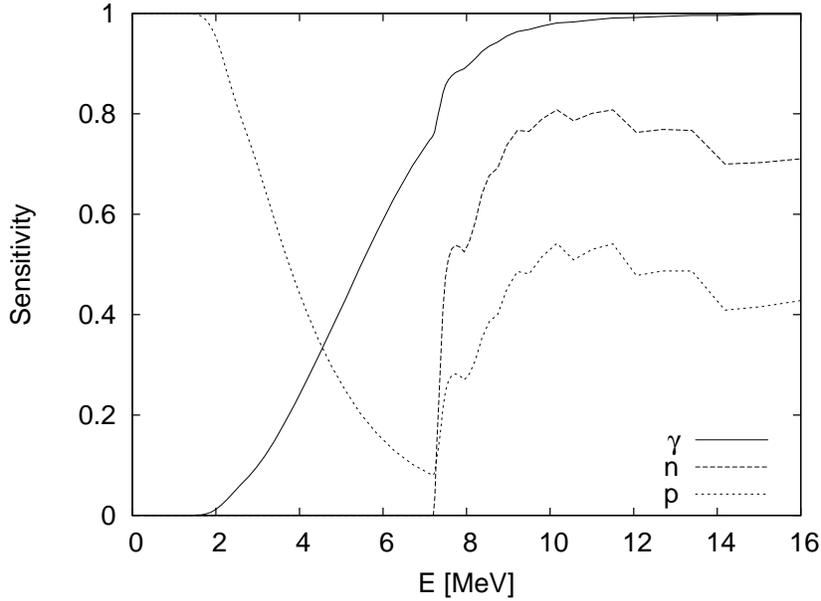}
  \caption{Sensitivities $s$ for $^{96}$Ru(p,$\gamma$)$^{97}$Rh}\label{fig:sensi}
\end{figure}

A helpful visual aid to estimate the relative importance of the different channels is a sensitivity plot. We define the sensitivity $s$ as a measure of a change in the cross section $f_\sigma=\sigma_\textrm{new}/\sigma_\textrm{old}$ as the result of a change in a width by the factor $f_\omega$, with $s=0$ when no change occurs and $s=1$ when the cross section changes by the same factor as the width:
\begin{equation}
s=\left\{ \begin{array}{cl}
\frac{f_\sigma-1}{f_\omega-1} & \mathrm{if}\, f_\sigma >1, f_\omega>1\, \mathrm{ or }\, f_\sigma<1, f_\omega<1\,, \\
\frac{1-f_\sigma}{(f_\omega-1)f_\sigma} & \mathrm{if }\, f_\sigma <1, f_\omega>1\, \mathrm{ or }\, f_\sigma>1, f_\omega<1\,.
\end{array} \right.
\end{equation}
Plotting $s$ as a function of the c.m.\ energy yields a plot like the example shown in Fig.\ \ref{fig:sensi}. Using this example for investigating the reaction $^{96}$Ru(p,$\gamma$)$^{97}$Rh, we find from \cite{tommyranges} that the Gamow window is $1.63 \leq E \leq 3.42$ MeV for the typical p-process temperature $T=2.5$ GK. It can clearly be seen in Fig.\ \ref{fig:sensi} that the sensitivities are very different at lower and higher energies. For example, a measurement closely below the neutron threshold would be in a region where $s$ is largest for the $\gamma$ width but smallest for the proton width, just the opposite of what is found in the astrophysically relevant energy region. Above the neutron threshold the situation is even more complicated because there is additional sensitivity to the neutron width (dominating $\Gamma_\mathrm{tot}$), although not as large as to $\Gamma_\gamma$. If any discrepancy between measured and predicted cross sections was found, it would be hard to disentangle the different contributions. In any case, no information on the astrophysically important proton width could be extracted from a measurement at the higher energies shown in Fig.\ \ref{fig:sensi}.

From the general considerations above it also follows that it is advantageous to use reactions with neutrons in one channel for investigating the sensitivity to the charged-particle optical potential, i.e., using ($\alpha$,n), (n,$\alpha$), (p,n), or (n,p) reactions. Except within a few keV above the neutron threshold, the neutron width will always be much larger than the charged-particle width, even at higher than astrophysical energies. Therefore it will cancel with the denominator in Eq.\ \eqref{eq:cs} and leave the pure energy dependence of the charged-particle width. On the other hand, information on the neutron potential is hard to get from reactions. But this also implies that the sensitivity of astrophysical rates to the neutron optical potential is not high.

On a side note, if averaged widths $\langle \Gamma \rangle$ or strength functions $\mathcal{S}$ can be determined experimentally, it is more useful to obtain the latter. They are directly proportional to the transmission coefficients $T$ calculated in the theoretical reaction models by solving the Schr\"odinger equation, without an additional dependence on the compound NLD.

\subsection{Stellar Effects}
\label{sec:sef}

\begin{table}
\begin{tabular}{llllll}
\hline
\tablehead{6}{c}{b}{Target: SEF} \\
\hline
$^{186}$W: $400$
& $^{185}$Re: $1300$
& $^{187}$Re: $1200$
& $^{190}$Pt: $5500$
& $^{192}$Pt: $3300$
& $^{198}$Pt: $310$ \\
 $^{197}$Au: $1100$
& $^{196}$Hg: $1700$
& $^{198}$Hg: $750$
& $^{204}$Hg: $43$
& $^{204}$Pb: $160$ & \\
\hline
\end{tabular}
\caption{Stellar enhancement factors (SEF) for selected ($\gamma$,n) reactions from NON-SMOKER calculations \cite{nonsmoker}}
\label{tab:sef}
\end{table}

Astrophysical investigations require the use of \textit{stellar} rates accounting for thermal population of excited states in the target. Only stellar rates allow to convert the rate for the forward reaction to the one for the reverse reaction by a simple relation derived from detailed balance. Stellar rates $r^*$ are calculated using stellar cross sections $\sigma^*=(\sum_\mu (2_\mu+1) \exp(-E_\mu/(kT) \sigma_\mu)/(\sum_\mu (2J_\mu+1) \exp(-E_\mu/(kT))$ (where $\mu$ indicates the excited target states) instead of laboratory cross sections $\sigma=\sigma_{\mu=0}$. It can be shown \cite{fowl,holm} that this leads to an equation identical to the one in Eq.\ \eqref{eq:integral} but with $\sigma$ replaced by an effective cross section $\sigma^\mathrm{eff}$, including sums over transitions to \emph{all} energetically accessible states in the exit \emph{and} entrance channel. This means that $\Gamma'_a$ in Eq.\ \eqref{eq:cs}, including only transitions to the target g.s., is replaced by a $\Gamma_a$ similar to $\Gamma_b$.

The stellar enhancement factor (SEF) $S=\sigma^*/\sigma$ is useful to assess the impact of the excited states. It is also useful to see how far the measured cross section is from the desired stellar cross section. A measurement should always be made in the direction of the SEF as close as possible to Unity, the stellar cross section (or rate) of the inverse reaction can then be calculated from detailed balance. From the definition of $\sigma^\mathrm{eff}$ and Eq.\ \eqref{eq:cs} it is easy to see that this will usually be the direction of positive reaction $Q$ value as in this case fewer transitions will contribute in the entrance channel than in the exit channel.
Photodisintegration reactions are the most extreme cases because they exhibit the largest difference in the number of contributing transitions between the entrance and exit channel. Table \ref{tab:sef} gives an example of the SEF appearing in such reactions \cite{nonsmoker,utsu}. Clearly, transitions from and to the g.s. (which are measured in the laboratory) contribute only a small fraction of the stellar cross section and it is better to measure capture reactions in order to be as close as possible to the stellar value. This is because the NLD is strongly increasing with excitation energy whereas the $\gamma$ strength function is decreasing with decreasing $\gamma$ energy. This leads to a maximal contribution to the $\gamma$ strength of $\gamma$ transitions $3-4$ MeV below the compound formation energy \cite{raugamma,litvi}.

There are some exceptions to the rule of measuring in the direction of positive $Q$ value, though \cite{coulett,coulsupp}. The range of transition energies in each channel varies from Zero to the particle separation energy. Transitions with a small relative energy are strongly suppressed by the presence of a barrier, e.g., the Coulomb barrier or the centrifugal barrier. If entrance and exit channel exhibit very different barriers, an endothermic reaction may show a smaller SEF than its inverse reaction although in principle the number of possible transitions is larger. Most of these transitions, however, may be suppressed by the barrier. The occurrence of this phenomenon depends on the barriers in the entrance and exit channels, and on the reaction $Q$ value. A very detailed explanation of the effect, along with a list of affected reactions, can be found in \cite{coulsupp}.

\section{Reaction Mechanisms}

\subsection{Modification of the statistical model}

The prediction of astrophysical reaction rates is difficult not just because the required nuclear properties are not fully known but also because several reaction mechanisms may contribute. Most astrophysically relevant reactions can be described in the HFM. Regular HFM calculations assume a compound formation probability independent of the compound NLD at the compound formation energy. Therefore the sum in Eq.\ \eqref{eq:cs} runs over all $J^\pi$ pairs (a high-spin cutoff is introduced in practical application of the model because spin values far removed from the spins appearing in the initial and final nuclei do not contribute to the widths). The availability of compound states and doorway states defines the applicability of the HFM \cite{rau97}. Relying on an average over resonances, the HFM is not applicable with a low NLD at compound formation. Single resonances and direct reactions will contribute then. On average the HFM will then overpredict the resonant cross section (unless single resonances dominate) because it will overestimate the compound formation probability. This can be treated by introducing a modification of the formation cross section which includes the compound NLD dependence. The summands of Eq.\ \eqref{eq:cs} will then be weighted according to the available number of states with the given $J^\pi$. (Formally this is the same as assuming $J^\pi$ dependent potentials for particle channels.) For an implementation of a parity dependence this was discussed in \cite{loens}. This was also used in the HFM code NON-SMOKER$^\mathrm{WEB}$ since version 4.0w \cite{websmoker}. Additionally, the option of weighting the HFM cross section by a NLD dependent function was offered. The new code SMARAGD \cite{smaragd,raureview,cyburt} has an improved version of this as default and uses $J^\pi$ dependent weighting of the summands, thus implicitly accounting for a low NLD at the compound formation energy. Preliminary results with this modification are shown in Fig.\ \ref{fig:modhf}. 

\subsection{Low-energy direct reactions}

\begin{figure}
\includegraphics[width=0.6\columnwidth,angle=-90]{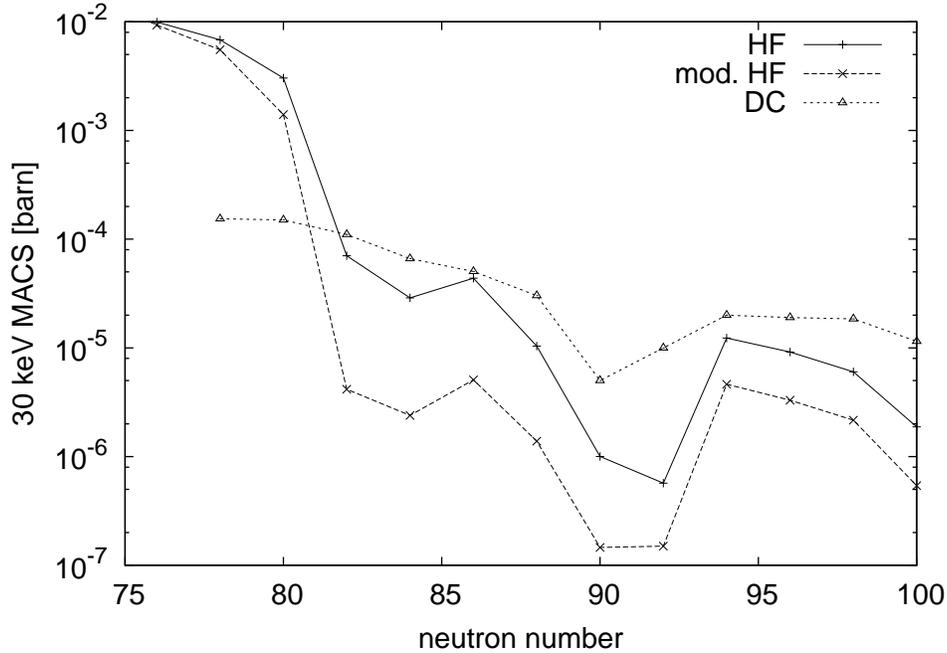}
\caption{\label{fig:modhf}Neutron capture cross sections of even Sn isotopes in an averaged DC model with energy-dependent spectroscopic factor (DC), a modified HFM (mod HF), and a standard HFM (HF) (preliminary results)}
\end{figure}

Due to the low particle separation energies and/or
NLDs encountered in very neutron- (or proton-) rich nuclei,
direct reactions will also become important, even at very low energies
(in the keV region) \cite{rau97,raujpg,descrau}. (The direct-semidirect mechanism at higher energies has been found to be of minor relevance for astrophysics \cite{bon07}.) Mainly relevant for astrophysical applications is direct capture (DC).
The relative importance of DC increases with decreasing neutron separation energy $S_\mathrm{n}$ for (n,$\gamma$) on neutron-rich nuclides \cite{raujpg} and with decreasing proton separation energy $S_\mathrm{p}$ for (p,$\gamma$) on proton-rich targets. This is because the number of final levels for $\gamma$ transitions from resonant capture becomes lower \cite{raugamma}. The Sn isotopes have implicitly a very low NLD and consequently the highest relative importance of DC. For this reason, they have become the focus of many experimental and theoretical investigations \cite{raugamma,litvi,loens,raujpg}.

The two major problems in the prediction of DC lie in the determination of the final states (i.e., the low-lying compound states) and their spectroscopic factors far off stability. Nuclear structure models cannot predict these states with sufficient accuracy and predicted DC cross sections differ by several orders of magnitude \cite{rau98}. Furthermore, most DC predictions far off stability have utilized only constant spectroscopic factors so far \cite{gordc97}, only few have employed energy-dependent ones \cite{raurep96,holz97,ejn98}. To circumvent the problem of the exact prediction of final states it has been suggested \cite{raurep96,holz97,gordc97} to employ averaged
properties, i.e.\ to replace the sum over discrete
final states by an integration over a level density, similar to the HFM. (The determination of the NLD poses the same problem as in the case of compound reactions.) 

The code SMARAGD also includes a global DC treatment using an averaged DC model and energy-dependent spectroscopic factors based on BCS and Lipkin-Nogami occupation probabilities \cite{bon07,raujpg,raureview,raujpg}. This average DC approach aims at providing robust predictions despite of considerable differences between microscopic predictions \cite{rau98}. Preliminary results for this DC treatment are shown in Fig.\ \ref{fig:modhf}.

The final rate (or cross section) is the sum of the modified HFM value and the DC one. Interestingly, for the Sn isotopes shown here (except for $N=92$) this sum is approximated by the unmodified HFM result within a factor of 10. This shows that it seems justified to use
unmodified HFM rates as crude estimate of the total rates for exotic nuclei.

This work was supported by the Swiss NSF, grant 200020-122287.





\bibliographystyle{aipproc}   

\end{document}